# The limits for complete photonic bandgaps in low-contrast media

Lukas Maiwald[1], Timo Sommer[2], Marvin Schulz[1], Manfred Eich[1,3], Alexander Yu. Petrov[1,4]*.

[1] *Institute of Optical and Electronic Materials, Hamburg University of Technology, Germany.*

[2] *QOLS, Blackett Laboratory, Imperial College London, London SW7 2AZ, United Kingdom.*

[3] *Institute of Materials Research, Helmholtz-Zentrum Geesthacht, Germany.*

[4] *ITMO University, St. Petersburg, Russia.*

*Correspondence to: a.petrov@tuhh.de.*

**Abstract:** The minimal refractive index contrast to obtain a complete photonic bandgap (CPBG) in structured media was not identified so far. We address this problem by considering distributed quasicrystals in with arbitrary number and positions of Bragg peaks in reciprocal space. For these structures an analytical estimation is derived which predicts that there is an optimal number of Bragg peaks for any refractive index contrast and finite CPBGs for an arbitrarily small refractive index contrast in 2D and 3D. Results of numerical simulations of dipole emission in 2D structures support our estimation. In 3D an emission suppression of almost 10 dB was demonstrated with a refractive index contrast of 1.6. The reason for residual leakage in 3D structures has to be further investigated.

Introduction

Photonic crystals (PhCs) have become an inherent part of photonics. The core property which allows them to function in the plethora of applications where they are used today is the blocking of light propagation for certain frequencies, i.e., their photonic bandgap (PBG) (*1–4*). However, another fundamental property of PhCs is their limited symmetry. This allows either the opening of a PBG only for certain directions at any small refractive index (RI) contrast or the opening of a complete PBG (CPBG), i.e., a PBG for all directions and all polarizations, only at a sufficiently high RI contrast (*5–7*). To achieve a CPBG, the PBGs opened for different directions of propagation should overlap which is achieved increasingly well the higher the symmetry of the structure is. For 2D structures the PBG should be opened for TE and TM polarization and the bandgap positions should coincide (*6*).

A CPBG has been realized in different classes of structures. In simple PhCs based on Bravais lattices the lowest refractive index contrast where a CPBG was found is $n_1/n_2 = 2.66/1 = 2.66$ for triangular 2D structures (*7*). In 3D structures based on a face-centered cubic lattice even an RI contrast of 3.5 was not sufficient (*8*). The optimization of the permittivity distribution in the unit cell of the PhC can be used to modify the strength of the Bragg peaks in reciprocal space and thereby to shape a more circular or spherical effective Brillouin zone (*9*). The minimal contrast numerically demonstrated for such a modified PhC in 2D is 1.16 for a TM bandgap in a holographic structure *(24)* and 2.1 for a CPBG in honeycomb-based structures (*6*). For 3D structures a minimal contrast of 1.9 was obtained numerically in an optimized diamond structure (*5*).

To overcome the symmetry limitations of PhCs, photonic quasicrystals have been investigated (*10–12*). Most quasicrystals are based on aperiodic tiling patterns, and 2D tilings with up to 14-fold symmetry have been presented (*13*). The realization of a CPBG in a 2D quasicrystal with 12-fold symmetry at a RI contrast down to 1.45 has been controversially discussed (*14–16*). It was later stated that such a 12-fold quasicrystal still requires a contrast of 2.65, which is only a marginal improvement over the 2D triangular PhC. On the other hand, a PBG for a single polarization was found at an extremely low contrast of approximately 1.1 in 10-fold (called 5-fold by the authors) and 12-fold quasicrystals (*12*). For 3D, quasicrystals based on an icosahedral tiling with 30 Bragg peaks have been shown (*17*). However, the strength of the corresponding 15 Bragg gratings was not sufficient to open up a CPBG with a RI contrast of 1.61 (*17*). Though localization was discussed later for a similar structure with RI contrast of 1.64 no CPBG was demonstrated (*18*). It should be mentioned that the quasicrystals considered so far all have many other Bragg peaks in reciprocal space corresponding to other periodicities. The strength of the Bragg peaks on a dedicated spherical surface was not optimized so far.

A CPBG can also be obtained in disordered structures without long range order at sufficient RI. A disordered arrangement of monodisperse spheres shows a suppression of the local density of states for RI contrasts down to 2.4 (*19*). Similarly, in hyperuniform disordered structures a CPBG with a contrast of 3.4 (*20*) as well as an omnidirectional TE bandgap at a contrast of 1.6 (*21*) have been shown.

Here we present an alternative approach to construct distributed 2D and 3D quasicrystals. These can have any number of Bragg peaks homogenously distributed over the whole angular range. Moreover, all Bragg peaks are at equal distances to the center of the reciprocal space. This way, the full strength of the RI contrast of the structure is concentrated in the defined Bragg gratings. The approach is similar to the dual-beam exposure technique (*22–24*) but with random phase shifts between different exposures in our case (*25, 26*). This randomization allows us to draw important conclusions about the connection between the total RI contrast and the RI modulation of single gratings. Based on that, we present an approximation for the obtained CPBG depending on the number of gratings and the available RI contrast. We find that there is an optimal number of gratings for a particular available RI contrast, and that a CPBG can be obtained for any, even arbitrarily small RI contrast. The predictions of the model are confirmed by numerical simulations in 2D. The trial structure in 3D shows significant emission suppression but it does not show a CPBG.

**Quasicrystal model**

The quasicrystal structures we propose are generated by a superposition of several 1D sinusoidal gratings which have their normal directions homogeneously distributed over the whole angular range. For the 2D case this can easily be realized by a uniform distribution over the azimuthal angle. 3D structures need a more complex distribution of gratings, which is described later. The superposition is mathematically described by

$$\Delta n_g(\mathbf{r}) = \sum_{i=1}^{N} \Delta n_i \cdot \sin(\mathbf{g_i} \cdot \mathbf{r} + \phi_i), \qquad (1)$$

where $N$ is the total number of gratings, $\Delta n_i$ is the RI amplitude of a single grating, $\mathbf{g_i}$ are the wave vectors defining the grating periods and directions and $\phi_i$ are the corresponding phases. Since the Fourier transform is a linear operation, a summation of 1D gratings corresponds to a summation of the Fourier transforms of each grating. In this case, the sine functions correspond to two Dirac delta functions, Bragg peaks, at $\pm \mathbf{g_i}$ in reciprocal space. The same periodicity, and thus, wave vector length $g = |\mathbf{g_i}|$ and amplitude $\Delta n_i$ is used for each grating but the phases $\phi_i$ are chosen randomly in order to obtain an isotropic structure. For a large number of gratings the local RI perturbation has a Gaussian distribution with standard deviation equal to $\Delta n_i \sqrt{N/2}$ (see Supplementary Text S1).

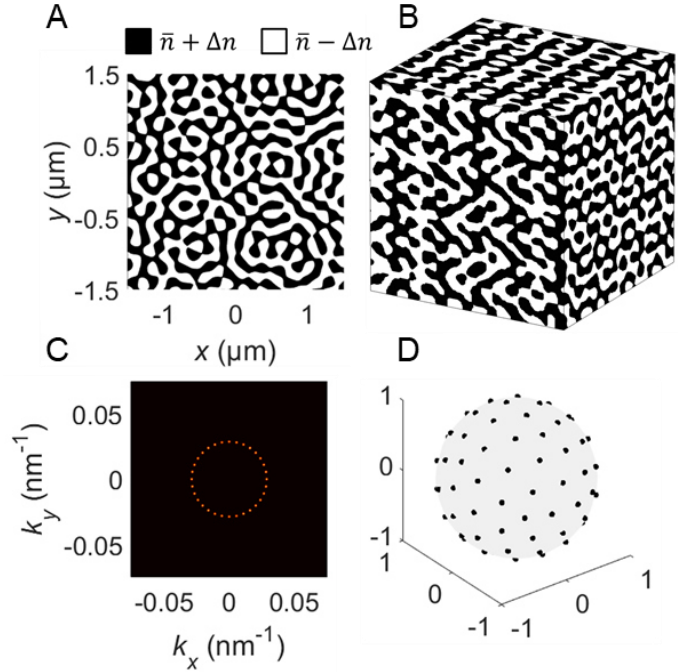

**Fig. 1**. Quasicrystals in real and reciprocal space. A, B: Examples of the investigated structures in 2D and 3D, respectively. The side lengths are 3 µm and a period of $g = 220$ nm was used in both cases. The 2D image was generated based on an overlap of 16 gratings; the 3D image is based on 46 gratings. C: Fourier transform of a circular excerpt of a 2D quasicrystal. A small diameter of the excerpt of 5 µm was chosen in order to make the Bragg peaks more visible. D: Schematic representation of the Bragg peaks used for the 3D structure.

To obtain a binary structure that can be represented by two materials the sum in Eq. (1) is then binarized by a sign function:

$$\Delta n_b(\mathbf{r}) = \Delta n \cdot \mathrm{sgn}(\Delta n_g(\mathbf{r})), \tag{2}$$

where $\Delta n$ is the refractive index perturbation from the average value $\bar{n}$, thus $n_1 = \bar{n} + \Delta n$ and $n_2 = \bar{n} - \Delta n$. Two examples of the structure in 2D and 3D are presented in Fig. 1. It can be shown that the binarized function still has approx. 64% of its intensity in the original gratings (see Supplementary Text S2). Each grating has an amplitude equal to $\Delta n_{i,b} = 2\Delta n / \sqrt{\pi N}$ (see

Supplementary Texts S1 and S2). The binarization also introduces noise in Fourier space that takes the residual approx. 36% of the intensity. However, no additional peaks other than the main ones and especially no higher harmonic orders are found in the reciprocal space. We therefore conclude that the large intensity content in the main Bragg peaks represents the optimal utilization of the RI contrast.

**CPBG estimation**

While every individual grating has a bandgap in its normal direction, other directions see an upshifted bandgap, corresponding to the Bragg condition. Thus, in the direction in-between two Bragg peaks the bandgap is at a slightly larger frequency. An omnidirectional bandgap for one polarization is achieved when the directional bandgaps have a sufficient opening to overlap. The effective Brillouin zone of the quasicrystal is schematically shown in Fig. 2A. We label the direction towards the Bragg peak of the grating with index $i$ as $\Gamma M_i$ and the direction in-between the two Bragg peaks with indices $i$ and $i+1$ as $\Gamma K_i$. Scanning the band diagram along the edge of the effective Brillouin zone the omnidirectional PBG opening can be evaluated (Fig. 2B). To achieve an omnidirectional PBG the upper edge of the PBG in $\Gamma M_i$-direction should be above the lower edge of the PBG in $\Gamma K_i$-direction.

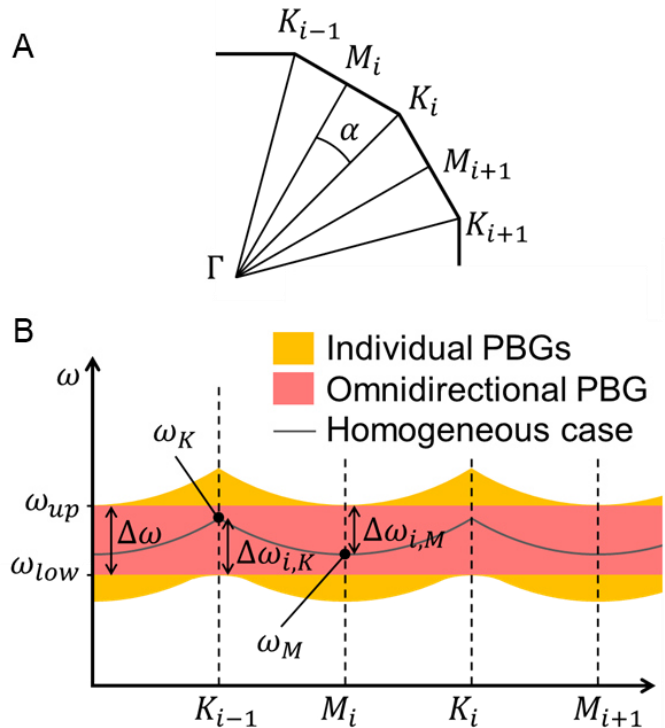

**Fig. 2.** A: Schematic representation of a section of the effective Brillouin zone of the 2D quasicrystal. B: Schematic band diagram showing the effective bandgaps in different directions. The bands depicted in gray show the band positions for the empty lattice case, when the grating contrast converges to zero. The orange shaded area represents the bandgap opening when a finite contrast is assumed. The red shaded area represents the omnidirectional PBG.

At the *M*-points we neglect the interaction with other gratings in the calculation of the directional bandgap of the single grating. At the *K*-points the neighboring gratings are contributing to the bandgap opening. Neglecting the influence of other gratings beyond the next neighbors, it can be shown that the PBG opening is by a factor $\sqrt{m}$ larger than for a single grating, where *m* is the number of interacting gratings (2 or 3 for 2D or 3D structures, respectively. See Supplementary Text S3). As shown in Supplementary Text S4, we end up with the following expression for the relative omnidirectional PBG opening:

$$\frac{\Delta\omega}{\omega} = \frac{\omega_{up} - \omega_{low}}{\omega} = \frac{\omega_M + \Delta\omega_{i,M} - (\omega_K - \Delta\omega_{i,K})}{\omega} = \left(\frac{1+\sqrt{m}}{2}\right)\sqrt{\frac{4}{\pi}\frac{\Delta n}{\bar{n}\sqrt{N}}} - \frac{\alpha^2}{2}, \quad (3)$$

where *α* is the angle between $\Gamma M_i$ and $\Gamma K_i$ directions. Then *α* can be expressed in terms of the number of gratings *N* to arrive at a function of the RI contrast and the grating number, only. However, the relation of these quantities is different for 2D and 3D.

The uniform distribution of the gratings in 2D is straight-forward and leads to the relation $\alpha = \pi/(2N)$. The resulting expression yields that the relative omnidirectional PBG opening reaches a maximum for a certain number of gratings:

$$N_{2D,opt} \approx 2.36\left(\frac{\bar{n}}{\Delta n}\right)^{\frac{2}{3}}; \quad \frac{\Delta\omega_{2D}}{\omega}(N_{2D,opt}) \approx 0.67\left(\frac{\Delta n}{\bar{n}}\right)^{\frac{4}{3}}. \quad (3)$$

Significantly, the bandgap opening will converge to zero as the contrast goes to zero, but for finite contrast values there will always be a finite omnidirectional PBG width.

To find a connection between the angle *α* and the grating number *N* in the 3D case, we make the approximation that each grating has exactly 6 neighbors at equal distance. In reality, the distribution closest to this approximation would be one with a Goldberg polyhedron as its effective Brillouin zone which would also have 12 pentagonal faces and differently sized hexagons (*27, 28*). Thus, our assumption slightly underestimates the maximal angle. Using this assumption, we obtain the connection $\alpha^2 = 4\pi/(3\sqrt{3}N)$ (see Supplementary Text S5). As in the 2D case, there is an optimum grating number and a corresponding optimum bandgap opening:

$$N_{3D,opt} \approx 2.46\left(\frac{\bar{n}}{\Delta n}\right)^{2}; \quad \frac{\Delta\omega_{3D}}{\omega}(N_{3D,opt}) \approx 0.49\left(\frac{\Delta n}{\bar{n}}\right)^{2}. \quad (4)$$

Again, the predicted bandgap persists even for a small but finite RI contrast.

In order for a PBG to be complete, it needs to inhibit propagation for all possible light polarizations. In 2D, the polarizations are fully described by an orthogonal basis of TE and TM polarizations. A shift between the bandgaps observed in TE and TM excitation is mainly caused by the different effective mean RI of the structure. For TM polarized light the E-field is always tangential to the material boundaries and therefore continuous. For TE polarized light the *E* and *D* fields can have all orientations towards the boundaries. However, in order to find the maximum difference between the effective permittivities for the two polarizations we may assume that all fields are normal to the boundaries. For that case, the relative difference in the bandgap positions for the different polarizations depends on the RI as (see Supplementary Text S6)

$$\frac{\Delta\omega_p}{\omega} = 2\left(\frac{\Delta n}{\bar{n}}\right)^2. \qquad (5)$$

In the 2D case the discrepancy between the bandgap positions for TM and TE polarization decreases faster than the maximal bandgap width for decreasing RI contrast. Thus, for low RI contrasts a better overlap of the bandgaps and thus a CPBG can be expected. For 3D structures the scaling power law with an exponent of 2 is the same and a better estimation is required to predict the existence of a CPBG. At the same time, the assumption that electric fields are only either parallel or orthogonal to interfaces represents an extreme case for 3D structures that will not be present in real field distributions. Therefore, in reality the birefringence will be smaller in the 3D structures and it might be that obtaining a CPBG is still possible.

**Simulation of the 2D case**

Simulations were done using the time domain solver of CST Studio Suite (*29*). A dipole emitter was placed in the center of the proposed structure. By orienting the dipole in the direction normal to the slab, TM modes are excited, while an orientation parallel to the slab leads to TE excitation. Accordingly, the vertical boundaries of the simulation volume must be terminated by perfect electric or magnetic conductor for TM or TE, respectively. The lateral directions are terminated by open boundaries acting as perfect absorbers. To probe the bandgap we evaluate the emitted power of the dipole $P$ by measuring the real part of its radiation resistance (*30–32*). The results are then normalized to the dipole emission into a homogeneous medium, $P_0$. A suppression of power emission is expected for frequencies inside the PBG. It is also expected that the emission of the dipole at these frequencies is decreasing exponentially with the lateral side length $L$ of the simulation volume (*19, 33*).

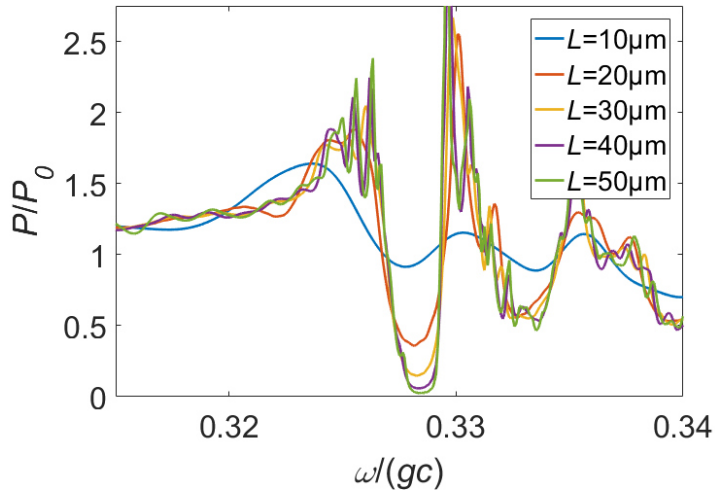

**Fig. 3.** Normalized power emission $P/P_0$ of a dipole for TM excitation placed inside a 2D quasicrystal with different structure sizes $L$. The quasicrystal has a RI contrast $n_1/n_2 = 1.58/1.42 \approx 1.11$ and $N = 16$ underlying gratings with period $g = 220$ nm.

The normalized power emission spectrum of the dipole for five different structure sizes with 16 gratings and a RI contrast of $n_1/n_2 = 1.58/1.42 \approx 1.11$ is shown in Fig. 3. An emission suppression band is seen in the spectra at a normalized frequency of about 0.328. The maximal suppression shows the expected exponential decay with increasing side length of the square-shaped simulated structure, unambiguously confirming the omnidirectional PBG properties (Fig. 4). As an example, the minima of the normalized power emission for different grating numbers $N$ and sizes of $L = 10$ to 50 µm are shown in Fig. 4.

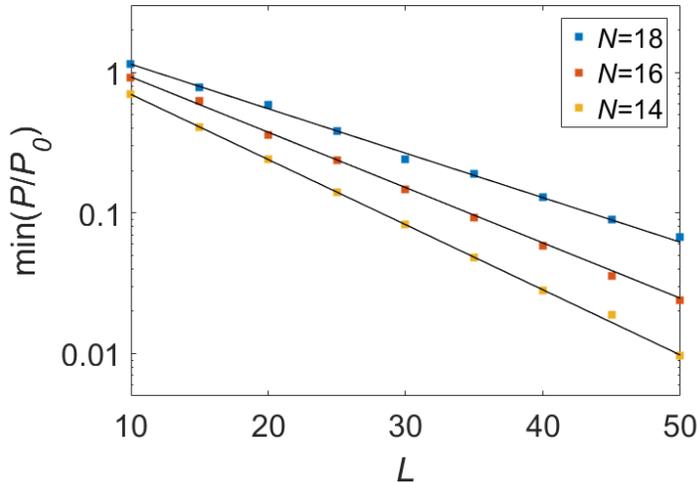

**Fig. 4.** Logarithmic plot of the minima of the normalized power emission over the edge length $L$ for a 2D structure with contrast $n_1/n_2 = 1.58/1.42$ and different numbers of underlying gratings. The black lines are linear fits.

In Fig. 5 we compare the PBG width predicted by the analytical model (Eq. (3)) to values obtained by simulation. We simulated grating numbers $N$ between 5 and 25. The bandgap width is obtained from the emission spectra since the width of the emission minimum remains approximately constant with increasing size $L$ of the simulated structure (see Supplementary Text S7). For some of the grating numbers the values for different structure realizations are shown. For a given $N$, these differ only in the seed of the random number generator defining the phases of the gratings. In the simulations the bandgap width is lower than the theoretically predicted width for most realizations. For large numbers of gratings, the omnidirectional PBG seems to disappear much faster than predicted. For these grating numbers the interaction between neighboring gratings cannot be neglected. Nevertheless, an omnidirectional PBG opening in the range of 0.5% is observed for grating numbers in the range of 10 to 20.

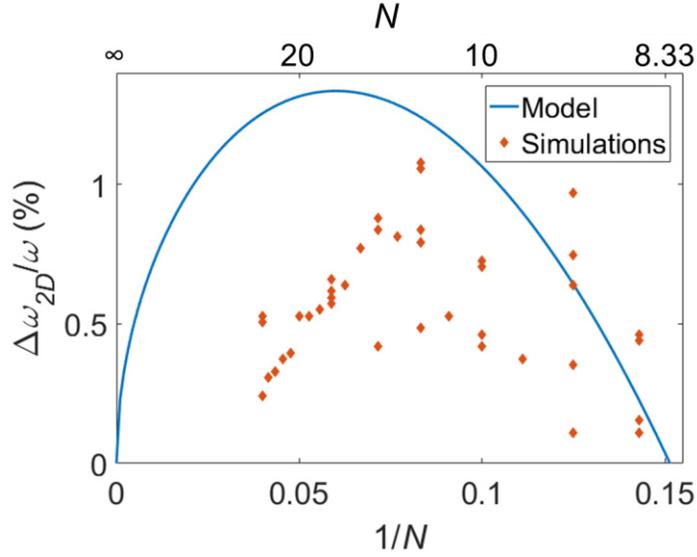

**Fig. 5.** Relative CPBG for 2D structure in TM polarization versus inverse number of gratings. Blue line – estimation, red diamonds – numerical simulations. The RI contrast is $n_1/n_2 = 1.58/1.42$. For simulations, a lateral structure size of $L = 50$ µm was used.

Additionally, for a CPBG the suppression experienced by different polarizations should coincide spectrally. While the previous results were obtained for TM excitation, we now orient the dipole and change the boundary conditions such that TE modes are excited. The comparison of the results for a grating number of $N = 16$ and an edge length of $L = 50$ μm is shown in Fig. 6. Although a slight spectral shift between the positions of the emission gap is observed there is a clear overlap region corresponding to a polarization independent CPBG. The overlap should further improve for smaller index contrast.

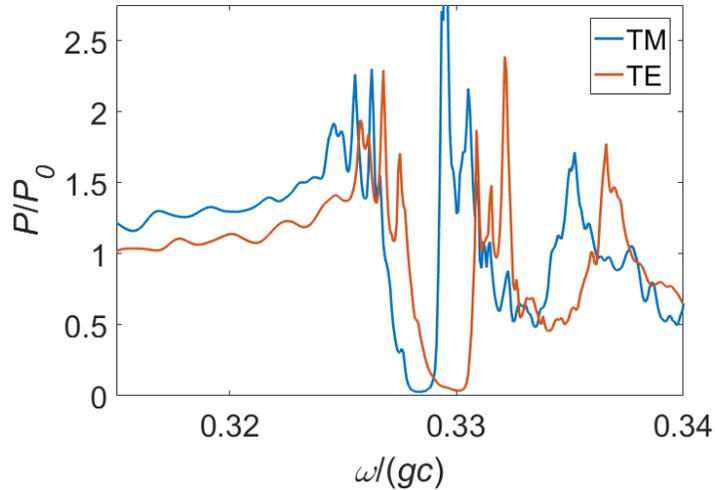

**Fig. 6.** Comparison of the normalized TM and TE power emission spectra of a dipole placed inside a 2D quasicrystal with contrast $n_1/n_2 = 1.58/1.42$, $N = 16$ underlying gratings and a lateral structure size of $L = 50$ μm.

**Simulation of the 3D case**

The gratings for the 3D structures need to be arranged such that the Bragg peaks are homogeneously distributed as points on the spherical surface with radius $g$ and the maximal distance from any point on the sphere to the closest Bragg peak is minimized. This task is a special type of a sphere covering problem where no exact solution exists for an arbitrary number of points (*28*). There are, however, solutions available in table form that are putatively optimal (*34*). Since each grating produces two Bragg peaks on opposite sides of the sphere a point symmetric distribution is necessary. The arrangements used for this work were the icosahedral solutions to the covering problem calculated in Ref. (*34*). For the results shown here the distribution of 92 points was used, corresponding to 46 gratings, the optimum as calculated by Eq. (4). The point distribution is listed in the Supplementary Materials, Table S1.

The simulation of low contrast 3D structures requires large simulation volumes. We have so far limited our consideration to structures with a RI contrast of $n_1/n_2 = 1.6/1$ and a maximum cube side length $L$ of 20 µm. The simulation yields a clear gap in the emission spectrum of the dipole (see Fig. 7). We did not observe an exponential decay of the emitted power over the side length $L$ and the emission suppression saturates to a nonzero value.

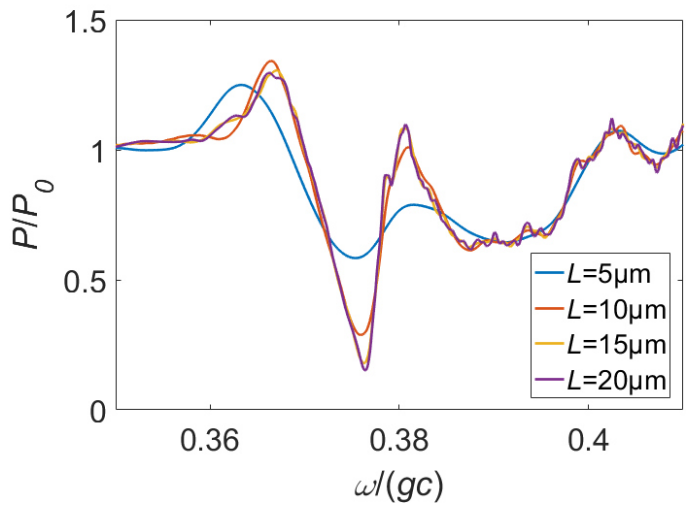

**Fig. 7.** Emitted power spectrum of a dipole placed inside a 3D quasicrystal with contrast $n_1/n_2 = 1.6/1$ and icosahedral distribution of the $N=46$ underlying gratings. The spectrum is normalized to the dipole emission in a homogeneous medium with $n=1.3$.

The fact that the theory predicts a CPBG which is not seen in simulations could be due to several reasons. Firstly, the chosen RI contrast might be too large for the analytical approximation to be applicable. Secondly, the interaction between gratings might no more be negligible. In this case, a smaller RI contrast and larger structures with more gratings may show a CPBG, but since increasing structure sizes are needed then, the 3D simulations become unfeasible. Thirdly, it might also be that the polarization effects are the limiting factor. According to the estimation the polarization effects can close the PBG as the bandgap opening and the polarization splitting both scale with the square of the RI contrast in 3D structures (Eq. (4) and (5)). The effect of polarization could be further studied by simulations using the scalar wave approximation and therefore eliminating the influence of polarization effects in the proposed structures.

In conclusion, we have proposed a distributed quasicrystal with an efficient utilization of the RI contrast to block emission in all directions. A simple analytical model is presented which predicts optimum conditions for the maximum bandgap opening for both the 2D and the 3D case, e.g., a bandgap opening for an arbitrarily small RI contrast. We show numerically that it is possible to obtain a 2D CPBG with a RI contrast as low as 1.11. This is the smallest RI contrast producing a CPBG that was demonstrated so far. The numerical confirmation for even smaller RI contrasts requires even larger simulation volumes but should be equally possible. The 2D structures shown in this work can readily be manufactured, e.g., by e-beam lithography or direct laser writing which paves the way to all kinds of photonic bandgap applications in slabs of materials with low RI contrasts. Even a typical RI contrast in the order of 1.1 (*35*) between the ordinary and extraordinary polarization of a typical liquid crystal would be sufficient.

In the 3D case we could not confirm the existence of a CPBG. Nevertheless, we demonstrate almost -10 dB suppression of emission at a contrast of 1.6 that could be realized with polymers. This by far exceeds the suppression shown in previous works investigating 3D structures at similar contrasts (*36, 37*).

30. A. E. Krasnok, A. P. Slobozhanyuk, C. R. Simovski, S. A. Tretyakov, A. N. Poddubny, A. E. Miroshnichenko, Y. S. Kivshar, P. A. Belov, An antenna model for the Purcell effect. *Sci. Rep.* **5**, 12956 (2015).

31. K. M. Schulz, H. Vu, S. Schwaiger, A. Rottler, T. Korn, D. Sonnenberg, T. Kipp, S. Mendach, Controlling the spontaneous emission rate of quantum wells in rolled-up hyperbolic metamaterials. *Phys. Rev. Lett.* **117**, 85503 (2016).

32. K. M. Schulz, D. Jalas, A. Y. Petrov, M. Eich, Reciprocity approach for calculating the Purcell effect for emission into an open optical system. *Opt. Express* **26**, 19247–19258 (2018).

33. A. Della Villa, S. Enoch, G. Tayeb, V. Pierro, V. Galdi, F. Capolino, Band gap formation and multiple scattering in photonic quasicrystals with a Penrose-type lattice. *Phys. Rev. Lett.* **94**, 183903 (2005).

34. R. H. Hardin, N. J. A. Sloane, W. D. Smith, Tables of spherical codes with icosahedral symmetry (available at http://neilsloane.com/icosahedral.codes/).

35. J. Li, C.-H. Wen, S. Gauza, R. Lu, S.-T. Wu, Refractive indices of liquid crystals for display applications. *J. Display Technol.* **1**, 51 (2005).

36. M. J. Ventura, M. Gu, Engineering spontaneous emission in a quantum-dot-doped polymer nanocomposite with three-dimensional photonic crystals. *Adv. Mater.* **20**, 1329–1332 (2008).

37. H. Yin, B. Dong, X. Liu, T. Zhan, L. Shi, J. Zi, E. Yablonovitch, Amorphous diamond-structured photonic crystal in the feather barbs of the scarlet macaw. *Proc. Natl. Acad. Sci. U.S.A.* **109**, 10798–10801 (2012).

38. The peak prominence is defined at https://www.mathworks.com/help/signal/ug/prominence.html.

**Acknowledgments:** We acknowledge the support from Dassault Systemes with their CST Studio Suite software. **Funding:** Deutsche Forschungsgemeinschaft (DFG) - Project number 278744289. **Authors contributions:** Conceptualization: A.P.; Data curation: L.M., T.S.; Formal analysis: L.M., T.S., A.P.; Funding acquisition: M.E., A.P.; Investigation: L.M., T.S.; Methodology: L.M., T.S., M.S., A.P.; Project administration: L.M., M.E., A.P.; Resources: L.M., M.E., A.P.; Software: L.M., T.S.; Supervision: M.E., A.P.; Validation: L.M., M.E., A.P.; Visualization: L.M., T.S.; Writing – original draft: L.M., T.S.; Writing – review & editing: L.M., T.S., M.S., M.E., A.P. **Competing interests:** The authors declare no competing interests. **Data and materials availability:** All data is available in the manuscript or the supplementary materials. Further information on the simulation implementation and data processing is available from the corresponding author.

**Supplementary Materials:**

Supplementary Text S1–S7

Figures S1–S5

Table S1

# Supplementary Materials for

## The limits for complete photonic bandgaps in low-contrast media

Lukas Maiwald, Timo Sommer, Marvin Schulz, Manfred Eich, Alexander Yu. Petrov

Correspondence to: a.petrov@tuhh.de

**This part includes:**

Supplementary Text S1–S7
Figures S1–S5
Table S1

# S1) Relation between single grating refractive index amplitude and the refractive index distribution of the sum before binarization

The quasicrystals before binarization are generated by a superposition of several sinusoidal gratings. At any point each of these gratings has the same refractive index (RI) amplitude $\Delta n_i$ and a random phase $\theta_j$. For a large number of gratings the RI of the resulting structure will, in each point, behave like the real part of a 2D random walk in the complex plane:

$$\Delta n_g = \sum_{j=1}^{N} \Delta n_i \text{Re}\{\exp(i\theta_j)\} = \Delta n_i \sum_{j=1}^{N} \cos(\theta_j), \tag{S1}$$

where $N$ is the total number of superposed gratings. The absolute square of this can be split into self-interacting and co-interacting grating terms.

$$\begin{aligned}
\left|\Delta n_g\right|^2 &= \Delta n_i^2 \sum_{j=1}^{N} \cos(\theta_j) \sum_{k=1}^{N} \cos(\theta_k) \\
&= \Delta n_i^2 \left( \sum_{\substack{j,k=1 \\ j=k}}^{N} \cos^2(\theta_j) + \sum_{\substack{j,k=1 \\ j \neq k}}^{N} \cos(\theta_j) \cos(\theta_k) \right)
\end{aligned} \tag{S2}$$

Averaging over this distribution, the first term has an expectation value of $N/2$ while the second term has an expectation value of zero, resulting in the standard deviation:

$$\sigma_g^2 = \langle |\Delta n_g|^2 \rangle = \Delta n_i^2 \frac{N}{2}. \tag{S3}$$

The relation between the individual grating RI amplitude and the average local RI perturbation is therefore given by the factor $\sqrt{N/2}$.

# S2) Influence of binarization on individual grating strength

The binarization that we apply to the quasicrystals is described by

$$\Delta n_b(\mathbf{r}) = \Delta n \cdot \text{sgn}\left(\Delta n_g(\mathbf{r})\right), \tag{S4}$$

Before the binarization step, all intensity in reciprocal space is in the Bragg peaks only. To find how much power is left in the Bragg peaks after binarization we calculate the correlation function of $\Delta n_b(\mathbf{r})$ and $\Delta n_g(\mathbf{r})$ over a large volume $V$:

$$C = \int_V \Delta n_g(\mathbf{r}) \Delta n_b(\mathbf{r}) d\mathbf{r} = \Delta n \int_V |\Delta n_g(\mathbf{r})| d\mathbf{r}. \tag{S5}$$

We may now choose to integrate over the possible values of the RI weighted by their probability $P(\Delta \tilde{n}_g)$ instead.

$$C = \Delta n \int_V |\Delta n_g(\mathbf{r})|\, d\mathbf{r} = V\Delta n \int_{-\infty}^{+\infty} |\Delta \tilde{n}_g| P(\Delta \tilde{n}_g)\, d\Delta \tilde{n}_g$$
$$= 2V\Delta n \int_0^{+\infty} \Delta \tilde{n}_g P(\Delta \tilde{n}_g)\, d\Delta \tilde{n}_g \tag{S6}$$

For a large number of gratings the RI is normally distributed with a mean value of zero and a standard deviation of $\sigma_g$, i.e.

$$P(\Delta \tilde{n}_g) = \frac{1}{\sigma_g \sqrt{2\pi}} \exp\left(-\frac{(\Delta \tilde{n}_g)^2}{2(\sigma_g)^2}\right). \tag{S7}$$

The integral in Eq. (S6) can then be solved and yields

$$C = \sqrt{\frac{2}{\pi}} \sigma_g \Delta n V. \tag{S8}$$

If we, in contrast, look at the self-correlation function of $\Delta n_g(\mathbf{r})$, we find that

$$SC = \int_V |\Delta n_g(\mathbf{r})|^2\, d\mathbf{r} = (\sigma_g)^2 V. \tag{S9}$$

We have a free choice of the grating amplitudes of the graded structure and we set $\sigma_g = \Delta n$. In this case the graded and binarized structures have the same self-correlation integral. Comparing Eq. (S9) and (S10) it can be seen that the binarization leads to a structure that has a $\sqrt{2/\pi} \approx 80\%$ similarity with the graded structure. Due to Parseval's theorem, a similar argument can be made in reciprocal space, where it becomes clear that this similarity can only be evaluated in the main Bragg peaks since the Fourier transform of the graded structure is otherwise zero. If we look at the square of the Fourier transforms instead, which is the relevant parameter to evaluate the scattering in the quasicrystal, we get $2/\pi \approx 64\%$ intensity inside the Bragg peaks of the binarized structure.

We looked at the Fourier transforms of actual structures that we numerically generated and could confirm this estimation. Fig. S1 shows the ratio of the Fourier transforms of a graded and a binary structure. The range around 0.8 is highlighted in red. We see that in all the peaks the ratio is inside the red range.

## S3) The directional bandgap opening in $\Gamma K$ direction

The frequency shift due to a small perturbation $\Delta \varepsilon$ in the permittivity can be described as *(4)*

$$\Delta \omega = -\frac{\omega}{2} \frac{\int \Delta \tilde{\varepsilon}(\mathbf{r}) \mathbf{E}(\mathbf{r}) \mathbf{E}^*(\mathbf{r})\, d^3\mathbf{r}}{\int \varepsilon(\mathbf{r}) \mathbf{E}(\mathbf{r}) \mathbf{E}^*(\mathbf{r})\, d^3\mathbf{r}}. \tag{S10}$$

We assume here a structure generated by the addition of only two neighboring gratings. The $\Delta\tilde{\varepsilon}$-term in Eq. (S10) can thus be written as

$$\Delta\tilde{\varepsilon} = \Delta\tilde{\varepsilon}_1 + \Delta\tilde{\varepsilon}_2 = \Delta\varepsilon_1 \cos(\mathbf{g}_1\mathbf{r} + \gamma_1) + \Delta\varepsilon_2 \cos(\mathbf{g}_2\mathbf{r} + \gamma_2). \tag{S11}$$

Additionally, we set the perturbation strength of all gratings to be equal, so $\Delta\varepsilon_1 = \Delta\varepsilon_2$. Further gratings at larger angles would primarily scatter at higher frequencies and are therefore omitted. Furthermore, we assume propagation in the direction that is furthest away from the normal grating directions, i.e., the $\Gamma K$ direction right in between two gratings. The Bragg condition is fulfilled for both gratings in this direction and the scattered light is propagating with wave vectors $\mathbf{k}_1$ and $\mathbf{k}_2$ as depicted in Fig. S2. The total electric field can now be described as a superposition of the different components,

$$\mathbf{E} = \mathbf{E}_0 \left( \exp(i\mathbf{k}\mathbf{r}) + \frac{1}{\sqrt{2}} \exp(i(\mathbf{k}_1\mathbf{r} + \phi_1)) + \frac{1}{\sqrt{2}} \exp(i(\mathbf{k}_2\mathbf{r} + \phi_2)) \right). \tag{S12}$$

The angle between the gratings is assumed to be very small. Therefore, the factors of $1/\sqrt{2}$ in the second and third term represent the fact that no power is transported at frequencies near the band-edge (zero group velocity).

The $\mathbf{EE}^*$-terms in Eq. (S10) can now be written as

$$\begin{aligned}\mathbf{EE}^* &= \mathbf{E}_0^2 \begin{pmatrix} 1 + \frac{1}{2} + \frac{1}{2} + \frac{1}{\sqrt{2}}\exp(i(\mathbf{k}\mathbf{r} - \mathbf{k}_1\mathbf{r} - \phi_1)) + \frac{1}{\sqrt{2}}\exp(i(\mathbf{k}\mathbf{r} - \mathbf{k}_2\mathbf{r} - \phi_2)) \\ + \frac{1}{\sqrt{2}}\exp(-i(\mathbf{k}\mathbf{r} - \mathbf{k}_1\mathbf{r} - \phi_1)) + \frac{1}{\sqrt{2}}\exp(-i(\mathbf{k}\mathbf{r} - \mathbf{k}_2\mathbf{r} - \phi_2)) \\ + \frac{1}{2}\exp(i(\mathbf{k}_1\mathbf{r} + \phi_1 - \mathbf{k}_2\mathbf{r} - \phi_2)) + \frac{1}{2}\exp(-i(\mathbf{k}_1\mathbf{r} + \phi_1 - \mathbf{k}_2\mathbf{r} - \phi_2)) \end{pmatrix} \\ &= \mathbf{E}_0^2 \begin{pmatrix} 2 + \sqrt{2}\cos(\mathbf{k}\mathbf{r} - \mathbf{k}_1\mathbf{r} - \phi_1) + \sqrt{2}\cos(\mathbf{k}\mathbf{r} - \mathbf{k}_2\mathbf{r} - \phi_2) \\ + \cos(\mathbf{k}_1\mathbf{r} - \mathbf{k}_2\mathbf{r} + \phi_1 - \phi_2) \end{pmatrix} \\ &= \mathbf{E}_0^2 \begin{pmatrix} 2 + \sqrt{2}\cos(\mathbf{g}_1\mathbf{r} + \phi_1) + \sqrt{2}\cos(\mathbf{g}_2\mathbf{r} + \phi_2) \\ + \cos((\mathbf{k}_1 - \mathbf{k}_2)\mathbf{r} + \phi_1 - \phi_2) \end{pmatrix},\end{aligned} \tag{S13}$$

where the vector relations depicted in Fig. S2 were used. Together with Eq. (S11) and using the fact that the field phase will be matched with the gratings ($\phi_m = \gamma_m$ for $m = 1,2$) we can write the integral in the numerator of Eq. (S10) as

$$\begin{aligned}&\int_V \Delta\tilde{\varepsilon} \mathbf{EE}^* \mathrm{d}^3\mathbf{r} \\ &= \Delta\varepsilon_1 \int_V \begin{pmatrix} 2\cos(\mathbf{g}_1\mathbf{r} + \phi_1) + 2\cos(\mathbf{g}_2\mathbf{r} + \phi_2) \\ +\sqrt{2}\cos^2(\mathbf{g}_1\mathbf{r} + \phi_1) + \sqrt{2}\cos(\mathbf{g}_1\mathbf{r} + \phi_1)\cos(\mathbf{g}_2\mathbf{r} + \phi_2) \\ +\sqrt{2}\cos^2(\mathbf{g}_2\mathbf{r} + \phi_2) + \sqrt{2}\cos(\mathbf{g}_2\mathbf{r} + \phi_2)\cos(\mathbf{g}_1\mathbf{r} + \phi_1) \\ +\cos(\mathbf{g}_1\mathbf{r} + \phi_1)\cos((\mathbf{k}_1 - \mathbf{k}_2)\mathbf{r} + \phi_1 - \phi_2) \\ +\cos(\mathbf{g}_2\mathbf{r} + \phi_2)\cos((\mathbf{k}_1 - \mathbf{k}_2)\mathbf{r} + \phi_1 - \phi_2) \end{pmatrix} \mathrm{d}^3\mathbf{r}.\end{aligned} \tag{S14}$$

When the integration is carried out over a large volume $V$, all terms in the integral except for the $\cos^2$-terms average out to zero. The result after integration is

$$\int_V \Delta\tilde{\varepsilon}\mathbf{E}\mathbf{E}^*d^3\mathbf{r} = \sqrt{2}\mathbf{E}_0^2 V\Delta\varepsilon_1. \tag{S15}$$

With the same argumentation we find the denominator to be

$$\int_V \varepsilon\mathbf{E}\mathbf{E}^*d^3\mathbf{r} = 2\mathbf{E}_0^2 V\varepsilon. \tag{S16}$$

Plugging this into Eq. (S10) yields

$$\frac{\Delta\omega_K}{\omega} = \frac{\sqrt{2}}{4}\frac{\Delta\varepsilon_1}{\varepsilon} = \frac{\sqrt{2}}{2}\frac{\Delta n_1}{n}. \tag{S17}$$

Note that the $\Delta\omega_K$ here just represents the one-sided shift of one of the bands. This deviates from the expression for the band shift from a single grating by the factor $\sqrt{2}$.

If we do the same calculation as above for a three dimensional structure where three instead of two gratings interact, we find a factor of $\sqrt{3}$ and the same considerations would hold true for higher dimensions. We therefore conclude, that the shift of the band due to the interaction of $m$ gratings is proportional to $\sqrt{m}$.

## S4) Derivation of the equation for the CPBG width

As shown in the main text, the frequency of the upper edge of the CPBG can be expressed as:

$$\omega_{up} = \omega_M + \frac{1}{2}\frac{\Delta n_{i,b}}{\bar{n}}\omega_M \tag{S18}$$

where $\omega_M = k_{\Gamma M}c/\bar{n}$ is the zero contrast band frequency in the M-point. Accordingly,

$$\omega_{low} = \omega_K - \frac{\sqrt{m}}{2}\frac{\Delta n_{i,b}}{\bar{n}}\omega_K \tag{S19}$$

where $\omega_K = k_{\Gamma K}c/\bar{n}$ is the zero contrast band frequency in the K-point. Now the absolute bandwidth of the CPBG can be expressed as

$$\Delta\omega = \omega_{up} - \omega_{low} = \omega_M + \frac{1}{2}\frac{\Delta n_{i,b}}{\bar{n}}\omega_M - \left(\omega_K - \frac{\sqrt{m}}{2}\frac{\Delta n_{i,b}}{\bar{n}}\omega_K\right) \tag{S20}$$

Applying the relation $k_{\Gamma K} = k_{\Gamma M}/\cos\alpha$ (see Fig. 2A in the main text) we obtain:

$$\frac{\Delta\omega}{\omega_M} = 1 + \frac{1}{2}\frac{\Delta n_{i,b}}{\bar{n}} - \frac{1}{\cos\alpha} + \frac{\sqrt{m}}{2}\frac{\Delta n_{i,b}}{\bar{n}}\frac{1}{\cos\alpha} \tag{S21}$$

For small $\alpha$ the cosine term can be expressed by the series expansion $1/\cos\alpha \approx 1 + \alpha^2/2$:

$$\frac{\Delta\omega}{\omega_M} = \left(\frac{1+\sqrt{m}}{2}\right)\frac{\Delta n_{i,b}}{\bar{n}} - \frac{\alpha^2}{2} + \frac{\sqrt{m}}{2}\frac{\Delta n_{i,b}}{\bar{n}}\frac{\alpha^2}{2}. \tag{S22}$$

If we now assume that many gratings are used for the generation of the quasicrystals, we may set $\omega_M \approx \omega$, where $\omega$ is the midgap frequency of the CPBG. The higher order last term in Eq. (S22) can be neglected:

$$\frac{\sqrt{m}}{2}\frac{\Delta n_{i,b}}{\bar{n}}\frac{\alpha^2}{2} \approx 0 \tag{S23}$$

With this and using $\Delta n_{i,b} = 2\Delta n/\sqrt{\pi N}$ as derived previously (A and B), the general expression as shown in the main text is obtained:

$$\frac{\Delta\omega}{\omega} = \left(\frac{1+\sqrt{m}}{2}\right)\sqrt{\frac{4}{\pi}\frac{\Delta n}{\bar{n}\sqrt{N}}} - \frac{\alpha^2}{2} \tag{S24}$$

We now need to treat the 2D and 3D case separately, as they show different relations between the angle $\alpha$ and the grating number $N$. In 2D a uniform distribution of the grating directions, and therefore of the Bragg peaks, over the half circle is straight-forward and leads to the relation $\alpha = \pi/(2N)$. Plugging this into Eq. (S24) yields an expression for the relative bandgap opening:

$$\frac{\Delta\omega_{2D}}{\omega} = \left(\frac{1+\sqrt{2}}{2}\right)\sqrt{\frac{4}{\pi}\frac{\Delta n}{\bar{n}\sqrt{N}}} - \frac{\pi^2}{8N^2}. \tag{S25}$$

This function has a maximum at a grating number of

$$N_{2D,opt} = \pi^{\frac{5}{3}}\left(\frac{1}{2+2\sqrt{2}}\right)^{\frac{2}{3}}\left(\frac{\bar{n}}{\Delta n}\right)^{\frac{2}{3}} \approx 2.359\left(\frac{\bar{n}}{\Delta n}\right)^{\frac{2}{3}}, \tag{S26}$$

and the bandgap opening at this grating number is

$$\frac{\Delta\omega_{2D}}{\omega}(N_{opt}) = \frac{3}{8}\left(\frac{2}{\pi}(1+\sqrt{2})\right)^{\frac{4}{3}}\left(\frac{\Delta n}{\bar{n}}\right)^{\frac{4}{3}} \approx 0.665\left(\frac{\Delta n}{\bar{n}}\right)^{\frac{4}{3}}. \tag{S27}$$

The relation between $\alpha$ and $N$ in 3D is explained in the next section. With the result (Eq. (S33)) we again obtain the expressions for the relative bandgap opening, the optimum grating number as well as the optimum bandgap opening:

$$\frac{\Delta\omega_{3D}}{\omega} = \left(\frac{1+\sqrt{3}}{2}\right)\sqrt{\frac{4}{\pi}\frac{\Delta n}{\bar{n}\sqrt{N}}} - \frac{2\pi}{3\sqrt{3}N}, \tag{S28}$$

$$N_{3D,opt} = \frac{1}{2+\sqrt{3}}\cdot\frac{8\pi^3}{27}\cdot\left(\frac{\bar{n}}{\Delta n}\right)^2 \approx 2.462\left(\frac{\bar{n}}{\Delta n}\right)^2, \tag{S29}$$

$$\frac{\Delta\omega_{3D}}{\omega}(N_{opt}) = \left(1+\frac{2}{\sqrt{3}}\right)\frac{9}{4\pi^2}\left(\frac{\Delta n}{\bar{n}}\right)^2 \approx 0.491\left(\frac{\Delta n}{\bar{n}}\right)^2. \tag{S30}$$

## S5) Distribution of gratings in 3D

The grating distribution in 3D is, just like the distribution in 2D, supposed to cover all available directions as evenly as possible in order to maximize the width of the complete photonic bandgap (CPBG). To achieve this, as discussed in the main text, we used an icosahedral distribution aiming for optimal coverage. The distribution is, however, a numerically found one and as such difficult to grasp analytically. Since we want to analytically relate the angular separation of grating directions to the number of gratings, we need an approximation for the angle distribution.

We used a simple approximation that tends to slightly underestimate the maximal angular separation of Bragg peaks by just assuming that the Bragg peaks are homogeneously distributed in a hexagonal pattern. In this case the effective Brillouin zone is approximated by a polyhedron with equal hexagonal facets. We wish to note that this is not a physically viable distribution as in real distributions at least twelve additional pentagons are needed to cover a sphere. Thus the maximal angular separation is slightly larger in a real situation. For a large number of gratings we disregard this deviation.

The solid angle of each hexagonal facet can be estimated for a large number of gratings as $4\pi$ divided by $2N$. Taking into account that the solid angle of one facet is (Fig. S3)

$$\Omega = \frac{3\sqrt{3}}{2}\alpha^2, \tag{S31}$$

the following equation is obtained:

$$4\pi = 2N\frac{3\sqrt{3}}{2}\alpha^2, \tag{S32}$$

and therefore

$$\alpha^2 = \frac{4\pi}{3\sqrt{3}N}. \tag{S33}$$

## S6) Refractive index difference for different polarizations

For TM polarized light the E-field is always tangential to the material boundaries and therefore continuous. The electric displacement $D$ however is averaged between the two different permittivities:

$$\bar{\varepsilon}_{TM} = \frac{\overline{D}}{\overline{E}} = \frac{\frac{1}{2}\varepsilon_1 E + \frac{1}{2}\varepsilon_2 E}{E} = \frac{\varepsilon_1 + \varepsilon_2}{2} = \varepsilon_1 + \Delta\varepsilon, \tag{S34}$$

where $\Delta\varepsilon = (\varepsilon_2 - \varepsilon_1)/2$.

For TE polarized light the $E$ and $D$ fields can have all orientations towards the boundaries. However, in order to calculate the maximum difference between the effective permittivities for the two polarizations we may assume that all fields are normal to the boundaries. For that case we obtain:

$$\bar{\varepsilon}_{TE} = \frac{\overline{D}}{\overline{E}} = \frac{D}{\frac{D}{2\varepsilon_1} + \frac{D}{2\varepsilon_2}} = \frac{2\varepsilon_1\varepsilon_2}{\varepsilon_1 + \varepsilon_2} = \frac{\varepsilon_1 + 2\Delta\varepsilon}{1 + \frac{\Delta\varepsilon}{\varepsilon_1}}, \tag{S35}$$

which can then be expressed into a Taylor series,

$$\bar{\varepsilon}_{TE} \approx (\varepsilon_1 + 2\Delta\varepsilon)\left(1 - \frac{\Delta\varepsilon}{\varepsilon_1} + \left(\frac{\Delta\varepsilon}{\varepsilon_1}\right)^2 - \cdots\right) = \varepsilon_1 + \Delta\varepsilon - \frac{(\Delta\varepsilon)^2}{\varepsilon_1}, \tag{S36}$$

where terms containing third or higher powers of $\Delta\varepsilon$ were neglected. The difference in effective permittivities for the two polarizations is thus found to be

$$\Delta\varepsilon_p = \bar{\varepsilon}_{TM} - \bar{\varepsilon}_{TE} = \frac{(\Delta\varepsilon)^2}{\varepsilon_1} \approx \frac{(\Delta\varepsilon)^2}{\varepsilon}. \tag{S37}$$

Since the difference is very small, $\Delta\varepsilon_p = 2\Delta n_p \bar{n}$ as well as $\Delta\varepsilon = 2\Delta n \bar{n}$ hold. The relative difference between the bandgap positions for the different polarizations is then found as

$$\frac{\Delta\omega_p}{\omega} = \frac{\Delta n_p}{\bar{n}} = \frac{\Delta\varepsilon_p}{2\bar{n}^2} = 2\left(\frac{\Delta n}{\bar{n}}\right)^2. \tag{S38}$$

## S7) Estimation of the bandgap size based on dipole emission spectra

The simulations yield the emission spectra of the dipole in which the bandgap is clearly seen. However, for Fig. 5 we needed a quantitative measure of the bandgap width. Since the spectra show some noise and the bandgaps sometimes contain spurious peaks we had to make decisions on which parts of the spectra belong to the bandgap and which do not. Fig. S4 shows some exemplary spectra.

We used following algorithm to determine the bandwidths. First, the global minimum of the curve is found. From this point, the algorithm looks for the maxima which are closest to the minimum. Maxima that do not exceed a certain prominence are neglected. The width of the band gap is then defined at half of the lower maximum. The peak prominence is the height of the peak with respect to the higher of the two neighboring minima. A MATLAB algorithm is applied here *(38)*. In case of the curve shown in the main text, the prominence limit was set to 0.07. Choosing different peak prominence values only leads to minor fluctuations in the overall picture while the trend of the points always resembles the curve found from the analytical theory. Fig. S5 shows the bandwidths determined by the algorithm for different peak prominence values.

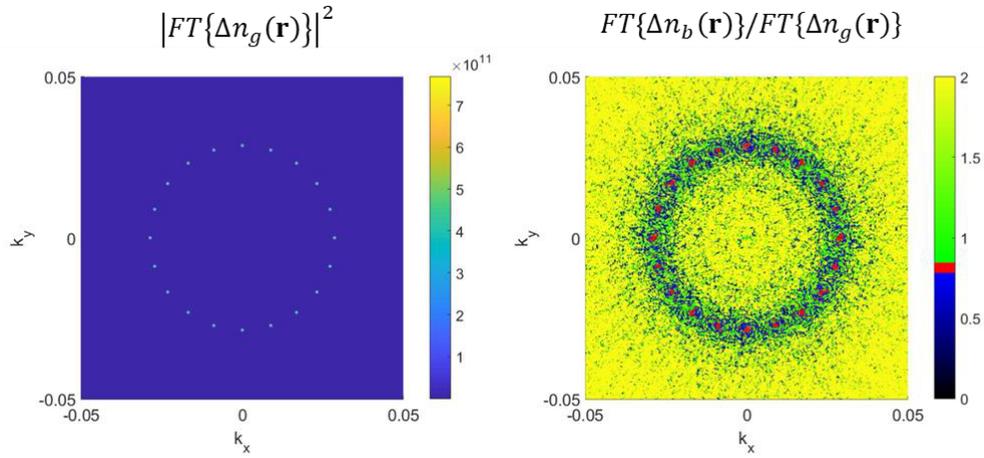

**Fig. S1.**
Left: Squared Fourier transform of the graded structure clearly showing the positions of the Bragg peaks. Right: Ratio of the Fourier transforms of the graded and the binarized structure. The value range of 0.766 to 0.828 is highlighted in red. The ratio in all the Bragg peaks lies in this range which is close to the expected $\sqrt{2/\pi}$ .

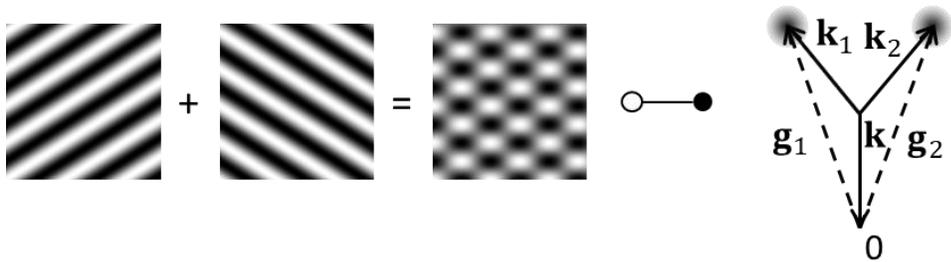

**Fig. S2.**
Two overlapping gratings with directions $\mathbf{g}_1$ and $\mathbf{g}_2$ in real (left) and reciprocal (right) space lead to scattering in the directions depicted as $\mathbf{k}_1, \mathbf{k}_2$.

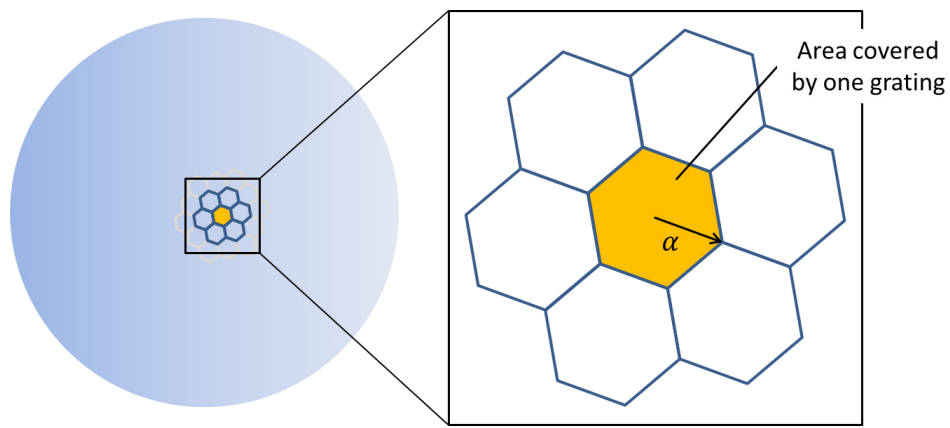

**Fig. S3.**
Schematic representation of the effective 3D Brillouin zone with only hexagonal facets.

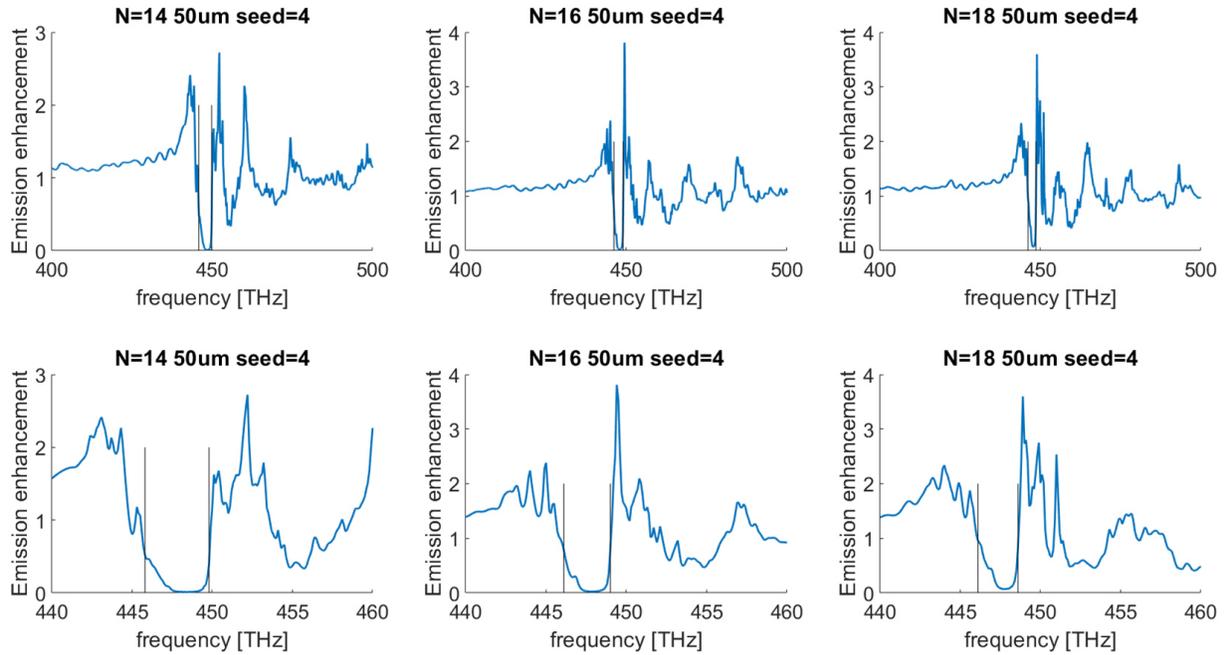

**Fig. S4.**

Emission spectra for grating numbers 14, 16 and 18. The figures in the bottom row are zooms into the figures from the top row. The black lines represent the frequencies that we assume to be the bandgap edges (see explanation in Supplementary Text G).

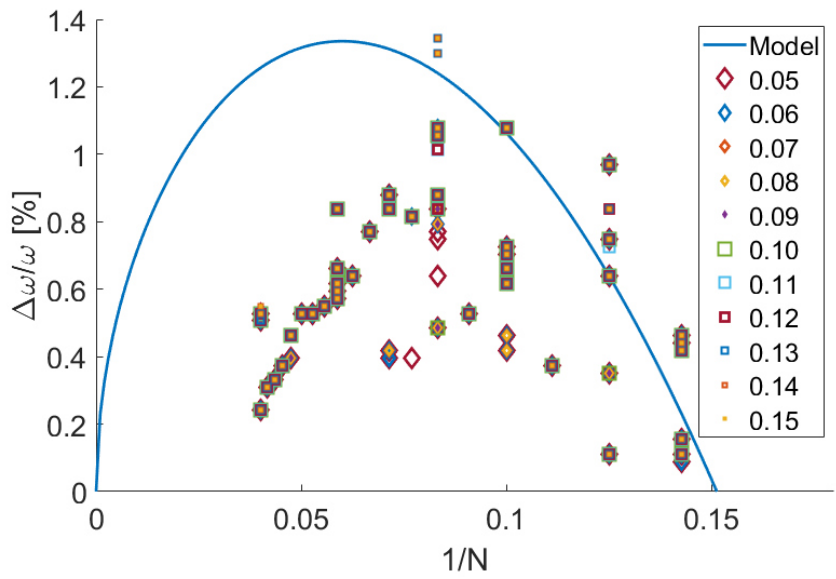

**Fig. S5.**
Reproduction of Fig. 5 from the main text with different choices of the peak prominence value.

| $k_x$ | $k_y$ | $k_z$ | $\phi$ |
|---|---|---|---|
| 0.403548212256 | 0.854728830047 | 0.326477361346 | 6,07602767609497 |
| 0.525731112119 | 0.850650808352 | 0 | 3,43836162763574 |
| 0 | 0.525731112119 | 0.850650808352 | 6,11155607905474 |
| 0.850650808352 | 0 | 0.525731112119 | 4,49132134879174 |
| -0.850650808352 | 0 | 0.525731112119 | 4,38395949910526 |
| 0 | -0.525731112119 | 0.850650808352 | 1,35773034366647 |
| 0.525731112119 | -0.850650808352 | 0 | 6,13411331002484 |
| 0.577350269190 | 0.577350269190 | 0.577350269190 | 0,0391458479614582 |
| -0.356822089773 | 0 | 0.934172358963 | 1,58953506230324 |
| 0.934172358963 | 0.356822089773 | 0 | 2,73187576808971 |
| -0.577350269190 | 0.577350269190 | 0.577350269190 | 4,89700732288120 |
| 0.577350269190 | -0.577350269190 | 0.577350269190 | 1,24209195617701 |
| 0 | 0.934172358963 | 0.356822089773 | 5,42234641811240 |
| -0.577350269190 | -0.577350269190 | 0.577350269190 | 6,17888868589838 |
| 0.934172358963 | -0.356822089773 | 0 | 1,02945116388937 |
| 0.356822089773 | 0 | 0.934172358963 | 3,75315985999858 |
| 0 | -0.93417235896 | 0.356822089773 | 0,0564613168336622 |
| -0.326477361546 | 0.403548212780 | 0.854728829723 | 2,42889900328402 |
| 0.730025573802 | 0.652954724054 | 0.201774106193 | 0,277465827159398 |
| -0.201774106587 | 0.979432085400 | 3.24e-10 | 6,01082787081183 |
| -0.201774105669 | 0.730025574126 | 0.652954723854 | 2,74039020345079 |
| 0.652954723854 | -0.201774105669 | 0.730025574126 | 5,96260027081677 |
| -0.979432085400 | 3.24e-10 | 0.201774106587 | 4,94050621777453 |
| 0.201774106193 | 0.730025573802 | 0.652954724054 | 5,44305519261544 |
| -0.326477361346 | -0.403548212256 | 0.854728830047 | 1,08803043205490 |
| 0.854728829723 | -0.326477361546 | 0.403548212780 | 0,470915860714625 |
| 0.326477361346 | 0.403548212256 | 0.854728830047 | 3,77457784035588 |
| 3.24e-10 | -0.201774106587 | 0.979432085400 | 1,05540035675435 |
| -3.24e-10 | 0.201774106587 | 0.979432085400 | 4,60796349307932 |
| -0.403548212780 | 0.854728829723 | 0.326477361546 | 2,56632846091485 |
| -0.854728830047 | -0.326477361346 | 0.403548212256 | 3,31694896282994 |
| -0.730025574126 | 0.652954723854 | 0.201774105669 | 5,89093600070965 |
| 0.652954724054 | 0.201774106193 | 0.730025573802 | 3,27791341135872 |
| 0.854728830047 | 0.326477361346 | 0.403548212256 | 0,679799073691103 |
| 0.403548212780 | -0.854728829723 | 0.326477361546 | 0,994146987275425 |
| -0.854728829723 | 0.326477361546 | 0.403548212780 | 3,42560929061698 |
| -0.201774106193 | -0.730025573802 | 0.652954724054 | 3,29492801091126 |
| -0.652954723854 | 0.201774105669 | 0.730025574126 | 4,00622331759963 |
| 0.201774105669 | -0.730025574126 | 0.652954723854 | 2,52267027206595 |
| -0.652954724054 | -0.201774106193 | 0.730025573802 | 4,08284591352056 |
| 0.201774106587 | -0.979432085400 | 3.24e-10 | 2,49379626665170 |
| 0.979432085400 | -3.24e-10 | 0.201774106587 | 3,92018055755879 |
| 0.326477361546 | -0.403548212780 | 0.854728829723 | 4,82174761387730 |
| -0.730025573802 | -0.652954724054 | 0.201774106193 | 1,12452624193891 |
| 0.730025574126 | -0.652954723854 | 0.201774105669 | 2,35981215092193 |
| -0.403548212256 | -0.854728830047 | 0.326477361346 | 3,15750833473500 |

**Table S1.**

The distribution of the grating directions used for the generation of the 3D quasicrystals was taken from *(34)*. The icosahedral distribution for optimal covering was used. Since each grating corresponds to two directions in reciprocal space we only needed half of the points. The exact k-vector coordinates normalized by $2\pi/g$ as well as the phases used for each grating are given in this table.